\documentclass[submission,copyright,creativecommons]{eptcs}
\providecommand{\event}{Post Proceedings of ADG'23} 
\usepackage[T1]{fontenc}
\usepackage{xurl,amsmath}
\usepackage{cite}
\usepackage{graphicx}
\usepackage{amsfonts}
\usepackage{color}
\usepackage{booktabs}
\usepackage{amssymb}
\usepackage{doi}
\usepackage{hyperref}

\begin{document}
\title{Using Java Geometry Expert as Guide \\in the Preparations for Math Contests}
\def\titlerunning{Using Java Geometry Expert as Guide in the Preparations for Math Contests}
\def\authorrunning{I.~Ganglmayr \& Z.~Kovács}
\author{
Ines Ganglmayr
\institute{The Private University College of Education of the Diocese of Linz, Austria}
\email{ines.ganglmayr@ph-linz.at}
\and
Zolt\'an Kov\'acs 
\institute{The Private University College of Education of the Diocese of Linz, Austria}
\email{zoltan.kovacs@ph-linz.at}
}
\maketitle
\begin{abstract}
We give an insight into Java Geometry Expert (JGEX) in use in a school context, focusing on the Austrian school system. JGEX can offer great support in some classroom situations, especially for solving mathematical competition tasks. Also, we discuss some limitations of the program.
\end{abstract}

\section{Introduction}

The use of technical media in Austrian mathematics lessons is largely limited to the GeoGebra medium. GeoGebra \cite{g1}
proved to be a great tool to visualize and analyze classroom problems, but certain tasks like proving geometric facts rigorously by using a visual explanation is not supported by GeoGebra. As an alternative approach, we focus on introducing JGEX \cite{j2}
as opposed to GeoGebra, specifically in the area of competition tasks.

Geometric proofs are no longer an important part of secondary school curriculum in Austria and many other countries. Formerly, however, Euclidean plane geometry and proving more complicated facts was a part of the expected knowledge of secondary level. There are, however, some initiatives, that call for rethinking school curriculum, by focusing on structured thinking again.

According to the Ministry of Education and Science in Portugal, for example, structured thinking is one of the main goals for teaching mathematics. It is also anchored in Austrian curricula that the qualitative development of tasks requires various dimensions of content, dimensions of action and dimensions of complexity, which hierarchically structure the understanding and learning of students. Analytically consistent thinking and the acquisition of mathematical skills are in the foreground. A central concern of mathematics is learning processes and techniques to acquire connections as well as insights and to solve problems \cite{P3}.\footnote{New version of the curriculum from Portugal: \url{https://eurydice.eacea.ec.europa.eu/national-education-systems/portugal/national-reforms-school-education}.}

One method to achieve this can be done by examining properties of certain structures. To achieve this, a concept of hierarchical structures is developed, and a systematic investigation is attempted. This form of working out is part of hypothetical-deductive thinking, which can be equated with mathematical thinking. But beyond that, inductive reasoning is also part of mathematical understanding, since it enables assumptions and conclusions \cite{P5}.

According to Duval, three cognitive processes are involved in learning geometry. These are visualization, construction and reasoning. Those can be activated separately but also together. They may or may not be interdependent. Indeed, Duval emphasizes that these three processes necessary for mathematical understanding and the mastery of geometry are closely related. JGEX can contribute to this and support children and young people in these three processes \cite{d6}.

In the context of a university course organized for prospective mathematics teachers at the Private University College of Education of the Diocese of Linz during the winter semester of 2022, some preliminary work was conducted to assess whether JGEX could be used to aid in solving geometric proof problems. This university course, under the guidance of the second author, primarily emphasized solving a broad spectrum of contest problems. As we will later demonstrate, JGEX proved to be an outstanding software tool for conducting such experiments. Additionally, participants in this university course attended a preparatory session for the Mathematics Olympiad in February 2023. During this session, they had the opportunity to gain firsthand experience and interact with young learners and their instructors, enabling them to derive conclusions regarding the feasibility of preparing for contest problems without technological assistance.

On the preparation day for the Mathematics Olympiad at Johannes Kepler University of Linz, Austria (JKU), 70 young learners geared up for a mathematics competition. The teaching staff focused on prime numbers and the pigeonhole principle as part of their preparations. The day commenced with an introduction to the content, followed by a 90-minute group exercise session facilitated by students \cite{P4}.

Even if the topic of prime numbers and the pigeonhole principle have not much to do with geometry, the tasks sometimes required a geometric overview and therefore structured thinking. The learners were sometimes overwhelmed with the processing of the geometric tasks. This observation was concluded by the participating prospective mathematics teachers, including the first author of this paper. A deficit in the three categories of visualization, construction and reasoning when solving the competition tasks with geometrical connections, was recognizable. The students often asked for help or verbally communicated their helplessness. While this is not yet covered with our research, it seems a plausible reason that structured thinking may require further support at all levels of school mathematics education.

In this paper and the related research, we assumed that JGEX can offer support in improving structured thinking. This geometry program may offer some missing steps in the visualization. Also, it can lead to constructions and above all it may provide various possibilities for arguments. Java Geometry Expert often guides students to solve more complex problems.

When solving competition tasks, the focus is on recognizing connections and grouping arguments for solutions. For this reason, JGEX is ideal in combination with competitive tasks.

We highlight that this paper is just a first report on our experiments, and it requires further research. However, we think that the first experiences are already promising.

\section{Testing of Competition Tasks Using JGEX}

JGEX is a complex software system that supports geometric proofs on “equational” properties of a planar figure. This means that ratios of quantities (like lengths of segments or size of angles) can be proven. Also, perpendicularity, parallelism, collinearity, concyclicity and related properties can be checked and proved. This limitation arises from the methods employed within JGEX, which encompass polynomial techniques such as Wu's method, the Gröbner basis method, as well as the Geometry Deductive Database method. Consequently, properties involving inequalities are not accommodated.

We show two examples. Below is a competition task that can be solved using JGEX and another task that is “equational” but JGEX does not provide any useful information on how to solve it. The tasks are preparatory materials for the mathematical competition in the field of geometry, collected by teacher Ralf Roupec from the Bundesrealgymnasium Freistadt, a well-known expert and team member for the preparations for math contests in Upper Austria.

\subsection{A Solvable Problem}

We consider a Problem of BAMO (Bay Area Mathematical Olympiad\footnote{The web page of the Bay Area Mathematical Olympiad is at \url{https://www.bamo.org}.}) 1999:

\begin{quotation}
Set $O=(0,0)$, $A=(0,a)$, $B=(0,b)$, where $0<a<b$. Let $k$ be the circle with diameter $AB$, and let $P$ be an arbitrary point on $k$. The line $PA$ intersects the $x$-axis at point $Q$.

Show that $\angle BQP=\angle BOP$.
\end{quotation}

\paragraph{A solution via JGEX.}

\begin{figure}[htbp!]
  \begin{center}
    \begin{tabular}{cp{10em}}
     \raisebox{-0.9\height}{\includegraphics[width=0.7\linewidth]{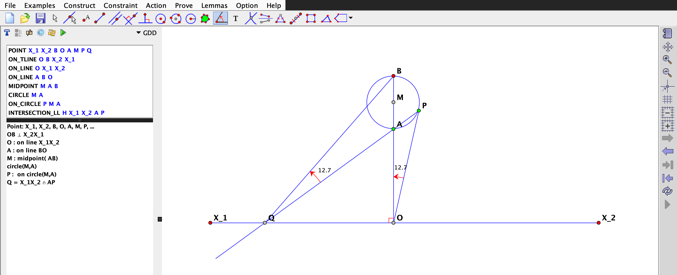}} & Draw the construction and then press \textit{Fix} (below).
    \end{tabular}
    \caption{A step-by-step solution provided by JGEX, step 1}
    \label{fig:step1}
  \end{center}
\end{figure}

\begin{figure}[htbp!]
  \begin{center}
    \begin{tabular}{cp{10em}}
     \raisebox{-0.9\height}{\includegraphics[width=0.7\linewidth]{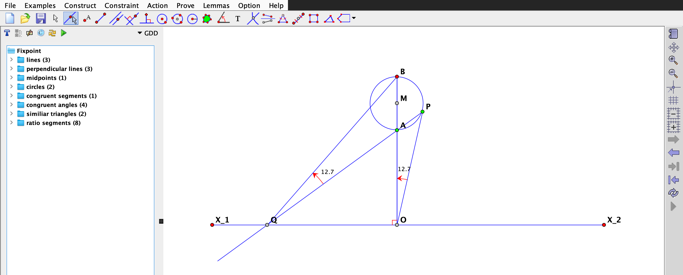}} & Open the folder: \textit{Congruent angles}.
    \end{tabular}
    \caption{A step-by-step solution provided by JGEX, step 2}
    \label{fig:step2}
  \end{center}
\end{figure}

\begin{figure}[htbp!]
  \begin{center}
    \begin{tabular}{cp{10em}}
     \raisebox{-0.9\height}{\includegraphics[width=0.7\linewidth]{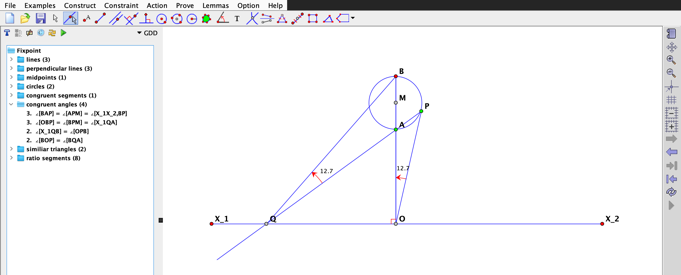}} &
    \end{tabular}
    \caption{A step-by-step solution provided by JGEX, step 3}
    \label{fig:step3}
  \end{center}
\end{figure}

\begin{figure}[htbp!]
  \begin{center}
    \begin{tabular}{cp{10em}}
     \raisebox{-0.9\height}{\includegraphics[width=0.7\linewidth]{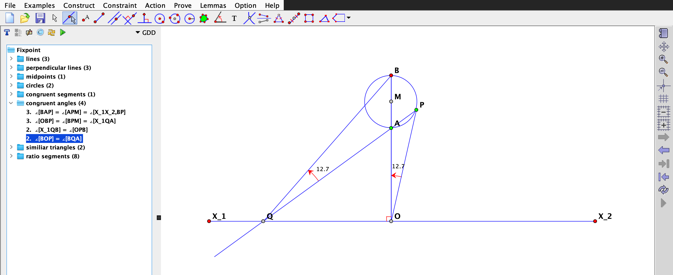}}
      & Select $\angle[BQP]=\angle[BQA]$ and then right click \textit{Prove}.
    \end{tabular}
    \caption{A step-by-step solution provided by JGEX, step 4}
    \label{fig:step4}
  \end{center}
\end{figure}

\begin{figure}[htbp!]
  \begin{center}
    \begin{tabular}{cp{10em}}
     \raisebox{-0.9\height}{\includegraphics[width=0.7\linewidth]{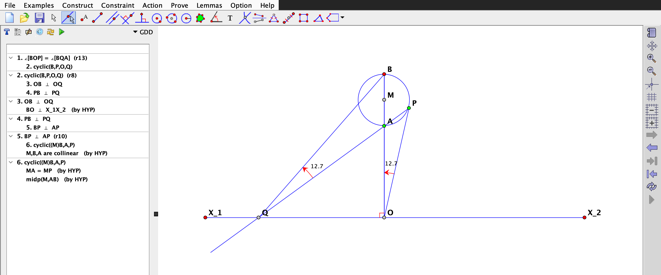}}
      & The steps of proof are shown.
    \end{tabular}
    \caption{A step-by-step solution provided by JGEX, step 5}
    \label{fig:step5}
  \end{center}
\end{figure}

Here JGEX supports step-by-step solving through important properties
that lead to the proof process. By collecting the various insights,
students are supported in solving the problem. See
figures~\ref{fig:step1}--\ref{fig:step5}. 

\begin{figure}[htbp!]
  \begin{center}
    \begin{tabular}{cp{10em}}
     \raisebox{-0.9\height}{\includegraphics[width=0.7\linewidth]{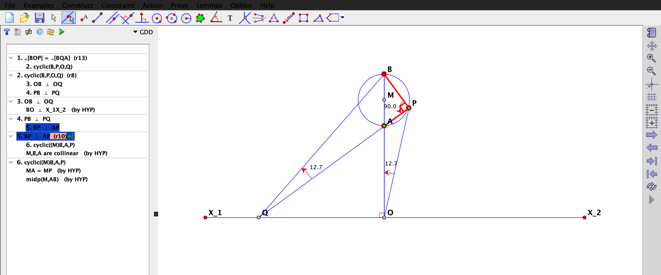}} & Observation: Right angle in triangle $BPA$.
    \end{tabular}
    \caption{Detailed information provided by JGEX, step 1}
    \label{fig:step6}
  \end{center}
\end{figure}

\begin{figure}[htbp!]
  \begin{center}
    \begin{tabular}{cp{10em}}
     \raisebox{-0.9\height}{\includegraphics[width=0.7\linewidth]{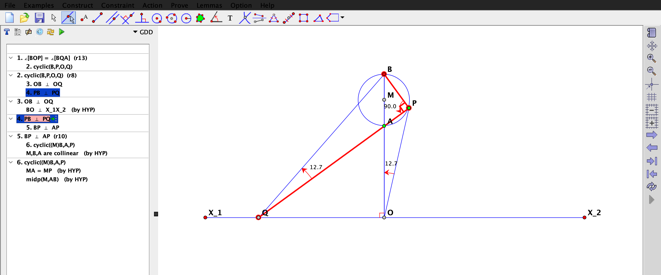}} & Observation: persistence of the right angle. (Right angle also in triangle $BQP$.)
    \end{tabular}
    \caption{Detailed information provided by JGEX, step 2}
    \label{fig:step7}
  \end{center}
\end{figure}

\begin{figure}[htbp!]
  \begin{center}
    \begin{tabular}{cp{10em}}
     \raisebox{-0.9\height}{\includegraphics[width=0.7\linewidth]{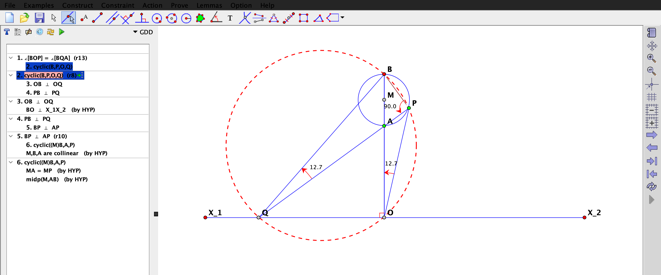}} & Observation: Circle through the points $B$, $P$, $Q$, $O$.

\textbf{Proof using the peripheral angle theorem}:

The red circle shows the existence of the peripheral angle theorem.

This is made possible by the chord $BP$ and concludes that the $\angle BQP=\angle BOP$.
    \end{tabular}
    \caption{Detailed information provided by JGEX, step 3}
    \label{fig:step8}
  \end{center}
\end{figure}

When performing further observations, JGEX can provide detailed information. See
figures~\ref{fig:step6}--\ref{fig:step8}.

\subsection{A Problem that is Difficult to Solve Automatically}

We consider another problem from Roupec’s collection:

\begin{quote}
Let $ABC$ be an acute triangle with circumcircle $k$. Let be $X$ the midpoint of the arc $BC$, that does not contain $A$. The points $Y$ and $Z$ are defined analogously.

Show that the orthocenter of $XYZ$ is center of the incircle of $ABC$.
\end{quote}

\begin{figure}[h]
\begin{center}
\includegraphics[width=0.3\textwidth]{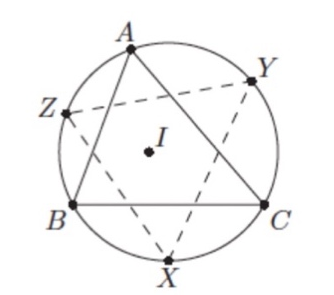}
\caption{A sketch of problem setting.}\label{fig:sketch}
\end{center}
\end{figure}

See figure \ref{fig:sketch} for a graphical explanation.

\paragraph{An attempt for getting the solution via JGEX.}

\begin{figure}[h]
\begin{center}
\includegraphics[width=0.8\textwidth]{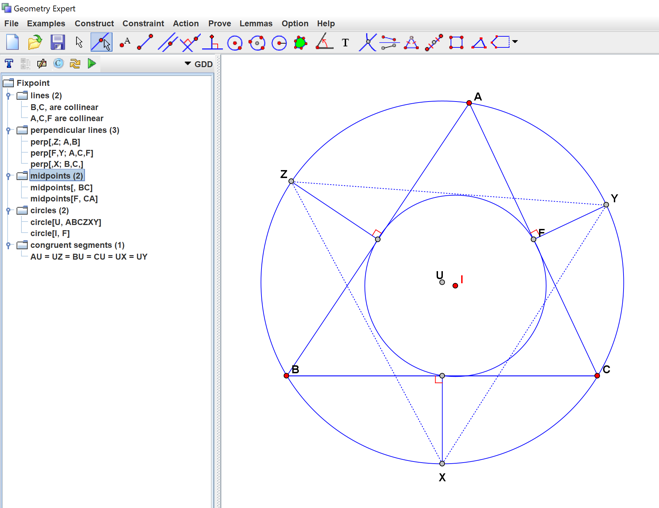}
\caption{An attempt to collect some fixed properties in JGEX.}\label{fig:difficult}
\end{center}
\end{figure}

\autoref{fig:difficult} shows that, after creating the figure in JGEX, there is a list provided by some fixed properties of the figure. There is, however, no automation provided to detect the property asked by the problem setting. One can try, however, other ways to formulate the problem, but an intuitive and quick way to obtain the required property (and its proof) does not seem to be accessible.

Here we highlight that JGEX provides several ways to construct the figure, we chose an intuitive one among the possible methods. For young learners, however, it may be challenging or even impossible to find the required formulation to achieve the expected property.

\section{Conclusion}
JGEX comes with a sophisticated user interface, however, the lack of support of languages (currently English, Chinese, German, Portuguese, Persian and Serbian are supported) and some other difficulties of its intuitive usage may make JGEX challenging to be used in classroom situations out-of-the-box. Among others, beginners may find difficult to look for the fixed properties of the given construction.

In our discussions during the university course activities, it turned out that JGEX can be indeed useful for the introduction of geometric topics in the classroom, including demonstrating the steps of visual proofs. After some technical introduction, students could construct simple figures and show connections with a mouse click.

Even though, it remains a challenging task to formulate the problem setting in a way that JGEX can provide a step-by-step proof. In our presentation we will show further examples in which JGEX can be of help, and in which it cannot provide a proof. For young learners, avoiding such limitations seems to be crucial for the everyday use.

An outlook for further research activities would be further work into how this program could be used in regular classes in a compliant manner.

\section{Acknowledgements}
The second author was partially
supported by a grant
PID2020-113192GB-I00 (Mathematical Visualization: Foundations, Algorithms
and Applications) from the Spanish MICINN.

\bibliographystyle{eptcs}
\bibliography{refs}

\end{document}